\newcommand{\al}{\mbox{$^{26}$\hspace{-0.2em}Al}}
\newcommand{\Msol}{M_{\sun}}

\newcommand{\gray}{\mbox{$\gamma$-ray}}

\newcommand{\pcmq}{\mbox{cm$^{-2}$}}

\newcommand{\psec}{\mbox{s$^{-1}$}}

\newcommand{\funit}{\mbox{ph \pcmq \psec}}
 
\newcommand{\frad}{\mbox{ph \pcmq \psec rad$^{-1}$}}

\def\phibar{\bar{\varphi}}

\def\MeV{\mbox{Me\hspace{-0.1em}V}}

\def\aj{AJ}		
	
\def\apj{ApJ}	
	
\def\apjs{ApJS}

\def\aap{A\&A}	
		
\def\aaps{A\&AS}

\documentclass{aa}
\usepackage{epsfig}
\begin{document}

\thesaurus{03(03.13.2; 03.20.1; 13.07.2; 02.14.1)}
\title{Image Reconstruction of COMPTEL 1.8 MeV \al\ Line Data}

\author{J.~Kn\"odlseder$^1$,
        D.~Dixon$^7$,
        K.~Bennett$^5$,
        H.~Bloemen$^3$,
        R.~Diehl$^2$,
        W.~Hermsen$^3$, 
        U.~Oberlack$^6$,
	    J.~Ryan$^4$, 
	    V.~Sch\"onfelder$^2$,
	    and 
	    P.~von Ballmoos$^1$}

\institute{$^1$Centre d'Etude Spatiale des Rayonnements, CNRS/UPS, B.P.~4346,
	       31028 Toulouse Cedex 4, France \\
           $^2$Max-Planck-Institut f\"ur extraterrestrische Physik,
	       Postfach 1603, 85740 Garching, Germany\\
	       $^3$SRON-Utrecht, Sorbonnelaan 2, 3584 CA Utrecht, The
	       Netherlands\\
	       $^4$Space Science Center, University of New Hampshire, Durham
	       NH 03824, U.S.A.\\
	       $^5$Astrophysics Division, ESTEC, ESA, 2200 AG
	       Noordwijk, The Netherlands \\
	       $^6$Astrophysics Laboratory, Columbia University, New York, 
	       NY 10027, U.S.A. \\
	       $^7$Institute for Geophysics and Planetary Physics, 
	       University of California, Riverside, CA 92521, U.S.A.}

\offprints{J\"urgen Kn\"odlseder}
\mail{knodlseder@cesr.fr}
\date{Received ; accepted }
\authorrunning{J. Kn\"odlseder et al.}
\titlerunning{Image Reconstruction of COMPTEL 1.8 MeV \al\ Line Data}
\maketitle

\begin{abstract}
We present a new algorithm, called Multiresolution Regularized 
Expectation Maximization (MREM), for the reconstruction of \gray\
intensity maps from COMPTEL data.
The algorithm is based on the iterative Richardson-Lucy scheme to 
which we added a wavelet thresholding step in order to eliminate 
image-noise in the reconstruction.
The wavelet thresholding explicitly accounts for spatial correlations 
in the data, and adapts the angular resolution locally, depending on 
the significance of the signal in the data.

We compare the performance of MREM to that of the maximum entropy and 
the Richardson-Lucy algorithms by means of Monte-Carlo simulations of 
COMPTEL 1.809 \MeV\ \gray\ line observations.
The simulations demonstrate that the maximum entropy and Richardson-Lucy 
algorithms provide virtually identical reconstructions which are heavily 
disturbed by image noise.
MREM largely suppresses this noise in the reconstructions, 
showing only the significant structures that are present in the data.

Application of MREM to COMPTEL 1.8 \MeV\ \gray\ line data results in a 
1.809 \MeV\ sky map that is much smoother than the maximum entropy or 
Richardson-Lucy reconstructions presented previously.
The essential features of this map are
(1) an asymmetric galactic ridge emission reaching from $l\approx45\degr$ 
to $l\approx240\degr$,
(2) a bright localised emission feature in the Cygnus region around
$(l,b)\approx(80\degr,0\degr)$,
(3) two emission spots at $l=317\degr$ and $l=332\degr$ situated in 
the galactic plane, and
(4) an extended emission region around $(l,b)\approx(160\degr,0\degr)$.
Comparison of the MREM map to the simulated reconstructions 
demonstrates that the 1.809 \MeV\ emission is confined to the 
galactic plane.

\end{abstract}

\section{Introduction}
\label{sec:intro}

Imaging the \gray\ sky at MeV energies by the COMPTEL telescope aboard 
the {\em Compton Gamma-Ray Observatory} ({\em CGRO}) presents a major 
methodological challenge.
Registered events are dominated by instrumental background, and 
additionally, source signals are widespread over the event parameter 
space.
Consequently, image recovery relies on a complex deconvolution 
procedure and on the accurate modelling of the instrumental background 
component.
A maximum entropy algorithm has been employed extensively for the 
reconstruction of intensity maps from COMPTEL data (Strong et al.~1992).
Recent examples of maximum entropy all-sky maps can be found in
Strong et al.~(1999) and Bloemen et al.~(1999a) for galactic continuum 
emission or in Oberlack et al.~(1996), Oberlack (1997) and Bloemen et 
al.~(1999b) for 1.809 \MeV\ \gray\ line radiation, attributed to the 
radioactive decay of \al.

Simulations revealed a tendency of clumpy reconstruction of emission 
in our maximum entropy images, leading to artificial `hot spots' of \gray\ 
emission in the reconstructions of diffuse emission distributions
(Kn\"odlseder et al.~1996).
From the images alone, these `hot spots' are indistinguishable from 
real point-like \gray\ sources, leading to considerable difficulties 
for the interpretation of the sky maps.
Indeed, the assessment of the significance of individual `hot spots' 
requires an substantial analysis effort, using simulations, Bootstrap 
analysis, and model fitting (e.g.~Oberlack 1997).

We understand the image lumpiness as the result of the weak constraints 
that are imposed on individual image pixels by our data.
COMPTEL images are usually reconstructed on a $1\degr\times1\degr$ pixel 
grid in order to exploit the telescope's angular location accuracy for 
point sources.
The fine pixelisation implies, however, that for weak diffuse emission, 
\gray\ intensities in individual pixels are generally not significant.
Increasing the pixel size could in principle avoid this problem
at the expense, however, of a reduced angular resolution.
We want to notice that this is not a particular property of the maximum 
entropy algorithm, but of every method that is operating on a fixed
grid of independent image pixels.
Apparently, significance and angular resolution are intimately 
related quantities (this relation is more generally known as the 
{\em bias-variance tradeoff}).

Algorithms that rely on a pre-defined pixel grid require an a priori 
choice of the angular resolution (by defining the pixel size) without 
constraining the significance of the fluxes in individual image pixels.
Alternatively, one may follow the opposite approach by choosing a 
priori the significance of image structures without constraining the 
angular resolution in the reconstruction.
An implementation of such an algorithm was discussed by
Pi\~na \& Puetter (1993) who introduced generalised image cells 
(called `pixons') to correlate adjacent image pixels according to the 
signal strength.
However, the application of Pixon based image reconstruction to 
COMPTEL 1.8 \MeV\ did not provide satisfactory results 
(Kn\"odlseder et al.~1996).

In this paper we present a new algorithm, called
{\em Multiresolution Regularized Expectation Maximization} (MREM),
which we developed in particular for the reconstruction of diffuse 
\gray\ emission.
We combined an expectation maximization (EM) algorithm with a 
multi-resolution analysis based on wavelets, which explicitly 
accounts for spatial correlations in the reconstructed image.
This leads to a convergent algorithm which automatically stops when 
the significant structure has been extracted from the data
(by significant structure we mean structure that will not change much 
under perturbation of the data).
The method requires an a priori choice of the significance level of 
emission structures while adapting the angular resolution according 
to the signal.

In the following we will present the MREM algorithm (\S 
\ref{sec:mrem}) and illustrate its performance by means of
simulations of COMPTEL observations (\S \ref{sec:simulations}).
The MREM algorithm is then applied for the reconstruction of an
1.809 \MeV\ all-sky map based on COMPTEL data obtained between May 1991 
to June 1996 (\S \ref{sec:comptel}).
This sky map will be compared to 1.8 \MeV\ all-sky maps presented 
previously which have been derived by the maximum entropy method 
(Oberlack et al.~1996) or the Richardson-Lucy algorithm 
(Kn\"odlseder et al.~1996).
A more theoretical description of the MREM algorithm will be given in 
a separate paper (Dixon et al., in preparation).

\section{The MREM algorithm}
\label{sec:mrem}

MREM is based on the Richardson-Lucy (RL) algorithm which has been 
proposed by Richardson (1972) and Lucy (1974) for the restoration of 
degraded images.
Given an initial estimate $f_j^0$ for the image, RL iteratively 
improves this estimate using
\begin{equation}
 f_j^{k+1} = f_j^k \left(
             \frac{\sum_{i=1}^N \frac{n_{i}}{e_{i}^k} R_{ij}}
                  {\sum_{i=1}^N R_{ij}} \right)
 \label{eq:rlmulti}
\end{equation}
($k$ denotes the iteration).
$R_{ij}$ is the instrumental response matrix which links the data 
space (indexed by $i$) to the image space (indexed by $j$).
For a given image $f_j^k$ and a given background model $b_i$, the 
expected number of counts in a data space cell is given by
$e_{i}^k = \sum_{j=1}^M R_{ij} f_j^k + b_{i}$.
The number of events observed in data space cell $i$ is given by
$n_{i}$.
It is easily seen that Eq.~(\ref{eq:rlmulti}) may be also written in 
the additive form
\begin{equation}
 f_j^{k+1} = f_j^k + \delta f_j^k ,
 \label{eq:rlstep}
\end{equation}
where
\begin{equation}
 \delta f_j^k = f_j^k \left(
                \frac{\sum_{i=1}^N \left( \frac{n_{i}}{e_{i}^k} - 1 
                      \right) R_{ij}}
                     {\sum_{i=1}^N R_{ij}} \right)
 \label{eq:rlcorrection}
\end{equation}
is the additive RL correction.

Shepp \& Vardi (1982) demonstrated that the Richardson-Lucy scheme is a 
special case of the expectation maximization (EM) algorithm 
(Dempster et al.~1977), and consequently it converges to the positively 
constrained maximum likelihood solution for Poisson data.
Due to the slow convergence of the algorithm, several modifications 
have been proposed to accelerate convergence
(Fessler \& Hero 1994).
For COMPTEL data we found that the ML-LINB-1 algorithm of Kaufman (1987) 
gives reasonable acceleration without degrading the reconstruction 
properties.
For ML-LINB-1, Eq.~(\ref{eq:rlstep}) is replaced by
\begin{equation}
 f_j^{k+1} = f_j^k + \lambda^k \delta f_j^k ,
 \label{eq:rlaccstep}
\end{equation}
where $\lambda^k$ is determined for each iteration using a line-search in 
order to maximise the likelihood for $f_j^{k+1}$ subject to the constraint 
$\lambda^k \delta f_j^k > -f_j^k$ 
(this constraint ensures the positivity of the intensities).

It is obvious from Eqs.~(\ref{eq:rlstep}) to (\ref{eq:rlaccstep}) 
that RL operates on a pre-defined pixel grid without any direct 
correlation between individual pixels.
In particular, apart from the convolution with the transpose of the 
response matrix $R_{ij}$, there is nothing which prevents RL from 
matching the estimates $e_{i}^k$ to the measurement $n_{i}$, and noise can 
easily propagate into the reconstruction where it is generally 
amplified.

For these reasons we added a multiresolution analysis to the iterative 
procedure which aims to correlate the image pixels and to extract only 
the significant structure from the data.

Each iteration of our MREM algorithm is composed of four steps:
First, we evaluate the normalised correction map
\begin{equation}
 \delta h_j^k = \frac{\sum_{i=1}^N \left( \frac{n_i}{e_i^k}-1 \right) R_{ij}}
                     {\sqrt{\sum_{i=1}^N \frac{R_{ij}^2}{e_i^k}}} ,
 \label{eq:mremstep1}
\end{equation}
for which $Var(\delta h_j^k)=1$.
Second, $\delta h_j^k$ is transformed into the wavelet domain where it is 
represented by a set of wavelet coefficients $w_m^l$,
$l$ representing the scales, and $m$ denoting the wavelet coefficients 
at this scale.
At scales $l > 1$ ($l=1$ represents a DC offset) the coefficients falling 
below a given threshold $\tau^l$ are zeroed by applying the operator
\begin{equation}
 \eta(w, \tau^l) = \left \{ \begin{array}{r@{\quad:\quad}l}
                            0 & |w| < \tau^l \\
                            w & |w| \ge \tau^l 
                            \end{array} \right. .
\end{equation}
This method is generally referred to as {\em wavelet thresholding} and 
has been proven successful for the removing of noise from a dataset 
without smoothing out sharp structures (Donoho 1993; Graps 1995).
Backtransformation of the nonzero coefficients from the wavelet 
domain into the image domain provides then a {\em de-noised} 
correction map $\delta \hat{h}_j^k$.
In compact matrix notation, the second step is given by
\begin{equation}
 \delta \hat{h}^k = {\bf W}^T  {\bf \eta} {\bf W} \delta h^k 
 \label{eq:mremstep2}
\end{equation}
where ${\bf W}$ is the discrete wavelet transform
(throughout this paper we use the translation invariant `cycle 
spinning' transformation of Coifman \& Donoho (1995) and employ 
Coiflet wavelets with 4 parameters).
Third, we calculate
\begin{equation}
 \delta f_j^k = f_j^k \delta \hat{h}_j^k \left( 
 \frac{\sqrt{\sum_{i=1}^N \frac{R_{ij}^2}{e_i^k}}}
      {\sum_{i=1}^N R_{ij}} \right) ,
 \label{eq:mremstep3}
\end{equation}
which, in absence of any wavelet thresholding, is equivalent to the 
original RL correction map Eq.~(\ref{eq:rlcorrection}).
In the last step, the previous estimate $f_j^k$ is updated using 
Eq.~(\ref{eq:rlaccstep}).
Due to the wavelet thresholding, positivity of the pixel intensities 
is not implicitly assured, and we explicitly require
$f_j^{k+1} \ge f_{\epsilon}$ where $f_{\epsilon}$ is a negligible 
intensity level.

For efficient de-noising, the scale-dependent thresholds $\tau^l$ have 
to be related to the expected statistical noise $\sigma^l$ in the 
wavelet domain at each scale.
We estimate $\sigma^l$ by simulations where we replace 
$n_i$ in Eq.~(\ref{eq:mremstep1}) by a Poisson derivate of $e_i^k$ 
and transformation of the resulting `mock correction map' into the wavelet 
domain.
In this approach it is important that the statistical noise in the MREM
correction map $\delta h_j^k$ is independent of the pixel location $j$.
For this reason we normalised $\delta h_j^k$ so that 
$Var(\delta h_j^k)=1$.
We define then $\tau^l = s \sigma^l$, where $s$ specifies the 
significance level below which structures should be suppressed in 
the reconstruction.
In the examples presented below we will vary $s$ between $2.5$ and 
$3.5$ in order to demonstrate the impact of the choice of $s$ on the 
reconstructed images.

The final critical aspect of MREM is that we do not calculate
corrections for all wavelet scales simultaneously.  
We begin by allowing in only corrections corresponding to the largest wavelet
scale, with the other scales simply being zeroed out; thus our estimate
in the initial iterations corresponds only to the average large
scale structure in the map.  
Once this converges, we then admit corrections from the next smallest 
wavelet scale, allow it to converge, and so forth.  
The resulting reconstruction is the final product of the algorithm, which 
generally requires between 20 and 40 iterations for our data.  

This procedure of progressively admitting smaller wavelets is crucial to 
the performance of MREM, and has some rather interesting ramifications 
which we shall discuss in detail in a subsequent paper (Dixon et al., 
in preparation).  
We briefly note here that its main purpose is to aid in the discrimination 
of noise-induced corrections from that structure which is ``stable'' in the
sense of being reproducible from different datasets.  
Examination of Eq.~(\ref{eq:rlcorrection}) indicates that if the estimates
$e_i^k$ are very different from the data $n_i$, the corresponding
corrections will also be large.  
Ideally, we would like this to occur {\em only} if the correction corresponds 
to some statistically interesting structure, but for arbitrary $e_i^k$ (e.g., 
the initial flat guess) this won't be the case.  
By first converging to a coarse approximation, we get an estimate that is 
``close'' to the next coarsest approximation, which generally forces the 
noise-induced corrections at that next smallest scale to be small compared 
to those which we deem ``interesting''.
It is further interesting to note that the unregularised RL iteration
tends to pick out the larger scale average structure first, only
adding details in later iterations, and in this sense the progressive
scale procedure dovetails nicely with the known characteristics of the
iteration.

The de-noising using the wavelet transform has the desired property 
of introducing pixel-to-pixel correlations in the image where the 
correlation length depends on the amount of structure in the data.
Regions of the sky with uniform emission will be represented by few 
large-scale wavelet coefficients, while point sources are represented 
by few small-scale coefficients.
An important feature of our algorithm is that it is convergent.
If all significant structure has been extracted from the data, where 
`significant' is defined by the choice of $s$, the thresholding
operator will zero all wavelet coefficients and consequently the correction 
map will be structureless.
At this point, further iterations won't alter the reconstructed image 
anymore, hence we stop the iterations.

\section{Simulations}
\label{sec:simulations}

To illustrate the performance of MREM with respect to the 
maximum entropy (ME) and the Richardson-Lucy (RL) algorithms, we apply 
them to simulated COMPTEL observations of 1.809 \MeV\ 
\gray\ line emission.
The mock data that are used in the simulations are based on a two-component 
data space model, composed of the instrumental background and adopted 
models for the \gray\ line distribution for two typical cases:
a smooth large-scale emission model, and a rather structured model 
with emission on many spatial scales.
The instrumental response and background were calculated as expected 
for the combination of observation periods $0.1-522.5$, corresponding to 
data taken between May 1991 to June 1996.
From both components of the data space model mock datasets were created
independently by means of a random number generator assuming Poisson noise.
Both components were then added and images have been reconstructed 
from the combined mock dataset.
For the reconstructions it has been assumed that the instrumental background 
is known precisely, hence the resulting images are not subject to 
possible systematic uncertainties of the employed background model.
They are sensitive, however, to statistical uncertainties which are 
due to the particular data `realisation' as obtained by the random 
sampling procedure.
To illustrate this sensitivity, the same mock dataset has been used for the 
instrumental background component in all simulations.

\begin{figure*}
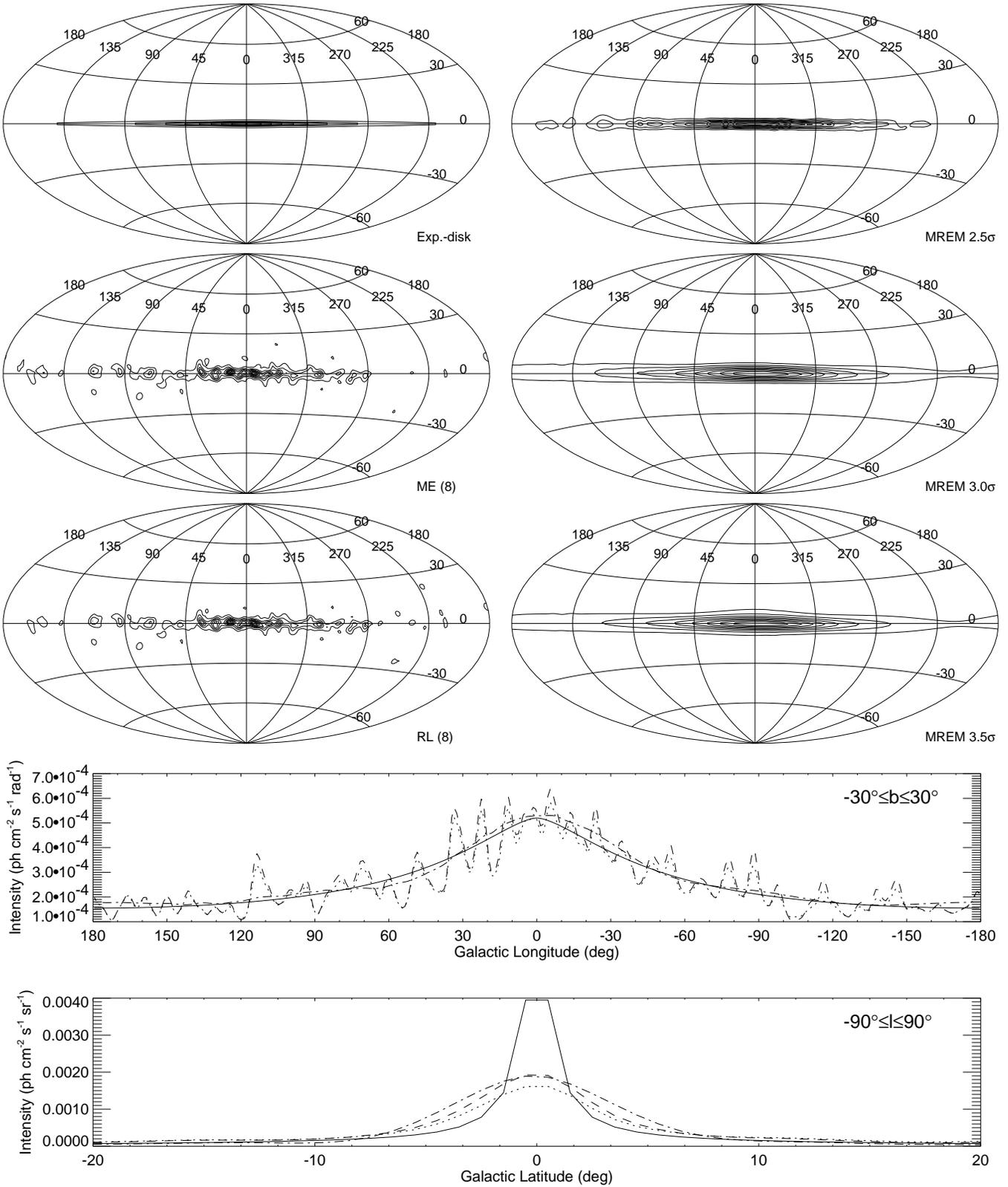

  \setlength{\unitlength}{1cm}
  \begin{minipage}[t]{18cm}
    \epsfxsize=8.9cm \epsfclipon
    \epsfbox{8399.f1a}
    \hfill
    \epsfxsize=8.9cm \epsfclipon
    \epsfbox{8399.f1d}
    \epsfxsize=8.9cm \epsfclipon
    \epsfbox{8399.f1b}
    \hfill
    \epsfxsize=8.9cm \epsfclipon
    \epsfbox{8399.f1e}
    \epsfxsize=8.9cm \epsfclipon
    \epsfbox{8399.f1c}
    \hfill
    \epsfxsize=8.9cm \epsfclipon
    \epsfbox{8399.f1f}
    \epsfxsize=18cm \epsfclipon
    \epsfbox{8399.f1g}
    \epsfxsize=18cm \epsfclipon
    \epsfbox{8399.f1h}    
    \caption{\label{fig:expsim}
    Reconstruction of an exponential disk intensity distribution 
    (Exp.-disk) using the 
    maximum entropy method (ME), the Richardson-Lucy algorithm (RL), 
    and the Multiresolution Regularized Expectation Maximization 
    algorithm (MREM).
    MREM reconstructions for the significance levels $s=2.5$, 
    $3.0$, and $3.5$ are shown.
    The bottom plots compare the longitude profile and latitude 
    profile of the reconstructions to that of the exponential 
    disk model; the model is shown as solid, the ME reconstruction 
    as dotted, the RL reconstruction as dashed, and the MREM ($s=3.0$) 
    reconstruction as dashed-dotted line.
    } 
  \end{minipage}
\end{figure*}

The following 1.809 \MeV\ model intensity distributions have been chosen.
First, we use an exponential disk model, i.e.~the intensity distribution 
that is expected if the galactic \al\ mass density would follow a double 
exponential law with scale radius of $R_0=4.5$ kpc and scale height of 
$z_0=90$ pc.
Model fitting has confirmed that these parameters provide a 
reasonable first-order description of 1.809 \MeV\ emission
(Kn\"odlseder 1997).
The total galactic \al\ mass has been normalised to $3\Msol$, a value 
slightly in excess from recent findings (e.g.~Diehl et al.~1997).
Second, the EGRET $>100$ MeV all-sky map was taken as template for 
the 1.809 \MeV\ intensity distribution.
The 1.809 \MeV\ intensity level of the map was adjusted to a plausible 
level by fitting the map to COMPTEL 1.8 \MeV\ \gray\ line data.
The first case testifies the response of the image reconstruction 
algorithms to a smooth intensity distribution while the second case 
represents probably a more realistic situation with structure on all 
spatial scales, from point-like to diffuse galactic plane emission.

\subsection{Exponential disk model}
\label{sec:expdisk}

The results of the exponential disk simulation are compiled in 
Fig.~\ref{fig:expsim}.
For comparison, the intensity distribution of the exponential disk 
model is also shown.
Since in our implementation ME and RL provide no criteria where to stop 
the iterative procedure, we used the correlation coefficient between the 
reconstruction and the model intensity map to determine the iteration 
which provides the smallest discrepancy to the model.
This is the case after iteration 8 for ME and the accelerated RL 
algorithm.

Both the ME and the RL reconstructions clearly pick up the emission 
ridge along the galactic plane with the highest intensities found 
towards the central radian ($-30\degr < l < 30\degr$).
The most striking difference between the model and the ME and RL
reconstructions, however, is the lumpiness of the recovered sky maps.
Although the emission follows in average the model intensity profile, 
it exhibits strong oscillations around this average, leading to `hot 
spots' and emission gaps along the galactic plane.
Indeed, these oscillations are already present in the first 
iterations of the reconstruction process and become more and more 
amplified with proceeding iterations.
If the iterations are pursued beyond those shown in Fig.~\ref{fig:expsim}, 
the oscillations will break up, and the image will be composed of nearly 
isolated point sources (Kn\"odlseder et al.~1996).

It is also interesting to recognise that the ME reconstruction is 
virtually identical to the RL reconstruction.
The difference between both algorithms is that ME imposes an 
additional constraint on the reconstructed image in that it `pushes' 
the image towards a `flat' sky map -- especially if the data 
are not very constraining.
This results in systematically lower fluxes for the ME 
reconstructions with respect to RL, which can be seen from the 
intensity profiles in Fig.~\ref{fig:expsim}.
If the ME iterations are proceeded further, and hence the entropy 
criterion is gradually weakened, the flux discrepancy between ME and 
RL disappears.
For this reason we always use `high' (e.g.~20-30) iterations when we 
determine fluxes from our ME sky maps.

\begin{figure*}
  \setlength{\unitlength}{1cm}
  \begin{minipage}[t]{18cm}
    \epsfxsize=18cm \epsfclipon
    \epsfbox{8399.f2}
    \caption{\label{fig:expres}
    Residual maximum likelihood ratio maps of the exponential disk
    simulations.
    Contour levels: $\sqrt{-2 \ln \lambda} = 2, 3, \ldots$. 
    From top to bottom the panels show 
    (a) the residuals of the background sample only (i.e.~the noise), 
    (b) the residuals of iteration 8 of the ME reconstruction,
    (c) the residuals of iteration 8 of the RL reconstruction,
    and (d) the residuals of the MREM ($s=3.0$) reconstruction.
    } 
  \end{minipage}
\end{figure*}

In contrast to ME and RL, the MREM reconstructions provide rather 
smooth emission distributions.
While some lumpiness remains for $s=2.5$, the images obtained 
with $s=3.0$ and $3.5$ show no `hot spots' or emission gaps.
In particular, the longitude profile is reasonably well reproduced 
and obeys only small deviations from the model distribution.
The most striking difference between the MREM sky maps and the 
model is the larger latitude extent of the reconstructions.
This, however, is not surprising since the width of the exponential 
disk model of $2.7\degr$ (FWHM) is considerably smaller than
the instrument's angular resolution of $4\degr$ (FWHM) at 1.8 \MeV\ 
(Sch\"onfelder et al.~1993).
Together with the weakness of the signal, this limits the achievable 
resolution in the reconstructions.
Indeed, the width of the latitude profile depends on the selected 
significance level $s$, rising from $5.3\degr$ for $s=2.5$ 
to $9.6\degr$ for $s=3.5$.
Obviously, the significance of the recovered emission features and 
the angular resolution are intimately related quantities.
Note that the width of the latitude profiles obtained by ME and RL 
is $5.7\degr$ (FWHM), which is also considerably wider than that of 
the model.

To judge the quality of the reconstructed images we determine the 
1.8 \MeV\ \gray\ line residuals by means of a maximum likelihood ratio 
test (de Boer et al.~1992).
For this purpose the sky maps of Fig.~\ref{fig:expsim} are convolved 
into the COMPTEL data space and added to the instrumental background 
model.
Residual emission is then searched by fitting point source models on 
top of the combined data space model for a grid of source positions.
The results of this point source search are shown in 
Fig.~\ref{fig:expres}. 
The quantity plotted is $-2 \ln \lambda$, where $\lambda$ is the maximum 
likelihood ratio $L(M)/L(S+M)$, $M$ represents the (two-component) data 
space model, and $S$ the source model which is moved over the sky area 
searched for residual emission.
In such a search, $-2 \ln \lambda$ obeys a $\chi^2_3$ distribution; in 
studies of a given source, $\chi^2_1$ applies.
In the latter case, the point source significance (in Gaussian 
$\sigma$) is given by $\sqrt{-2 \ln \lambda}$.

The top panel in Fig.~\ref{fig:expres} shows the residuals of the 
instrumental background sample only, hence reflects the statistical 
noise in the mock datasets (due to the dominance of the instrumental 
background component, the statistical noise is dominated by the 
background fluctuations).
In the ideal case, the residuals of the reconstructed images should 
be almost identical to those of the background sample.
Indeed, the residuals found on top of the MREM ($s=3.0$) 
reconstruction (panel d) are very similar to those expected for an 
ideal reconstruction (panel a).
The features are basically identical; only small deviations are found 
in their amplitude, e.g.~at $l\approx110\degr$ where MREM slightly 
overestimates the emission.
The ME reconstruction (panel b) forces image flatness, hence the 
residuals reflect a prominent flux suppression of the entire plane 
emission.
We therefore cannot easily detect to which extent noise has been 
included in the ME reconstruction.
For RL (panel c), no residual 1.8 \MeV\ emission is seen that correlates 
with the galactic plane.
In contrary, the likelihood ratios are even too small for the RL 
reconstruction with respect to the noise simulation, as expected if 
the data were overfit by the intensity map.
Additionally, prominent background features, like those at 
$(l,b)=(113\degr,2\degr)$ or at $(l,b)=(81\degr,-16\degr)$, are 
drastically reduced in both the ME and RL residual maps, yet are 
perceptible in the reconstructed intensity maps 
(cf.~Fig.~\ref{fig:expsim}).
This illustrates that statistical noise in the data is at least 
partially fit by the ME and RL sky maps.
It follows that the lumpiness of the ME and RL reconstructions are 
due to overfitting of the data.

\subsection{EGRET $>100$ MeV sky map}
\label{sec:egret}

\begin{figure*}
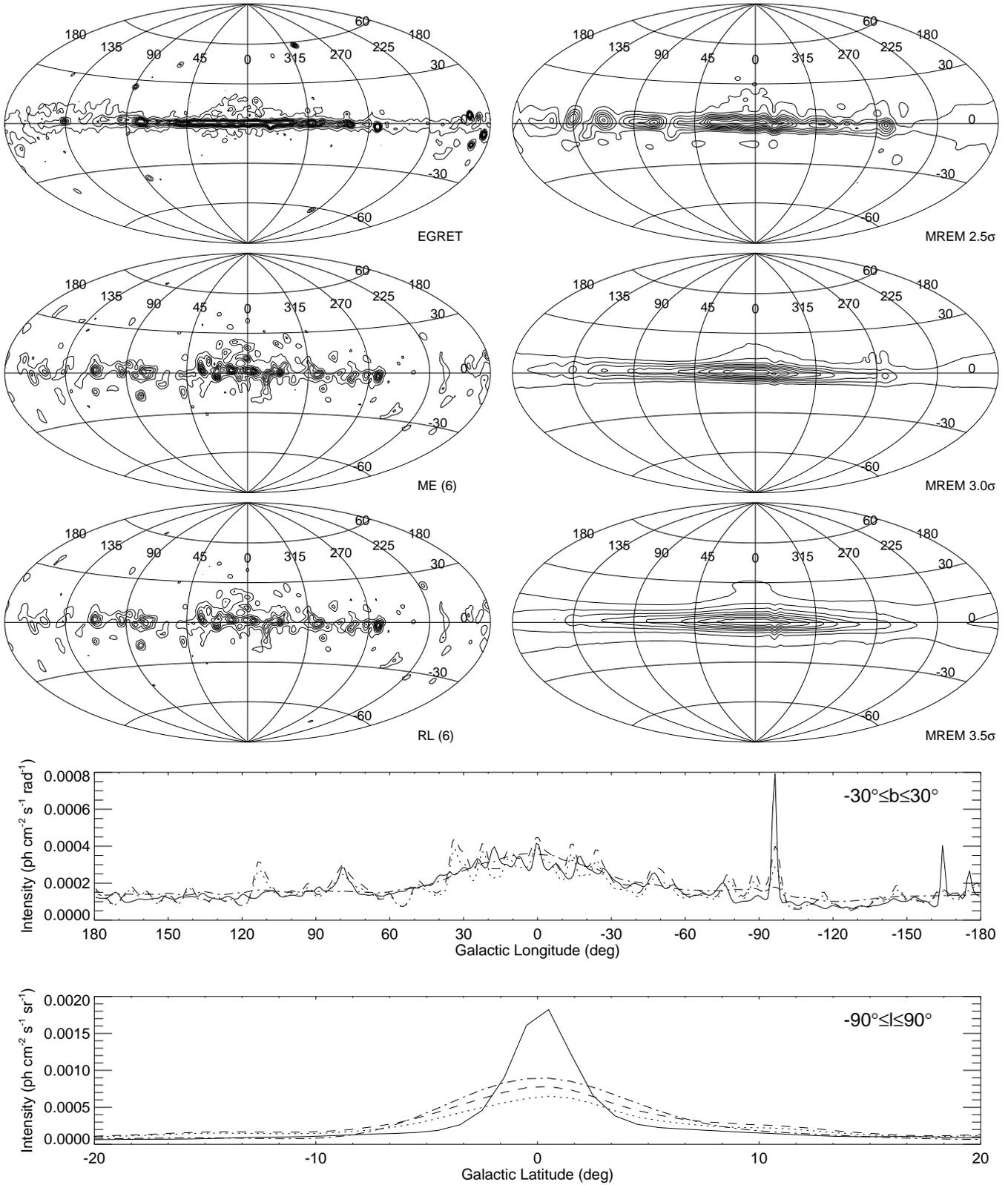

  \setlength{\unitlength}{1cm}
  \begin{minipage}[t]{18cm}
    \epsfxsize=8.9cm \epsfclipon
    \epsfbox{8399.f3a}
    \hfill
    \epsfxsize=8.9cm \epsfclipon
    \epsfbox{8399.f3d}
    \epsfxsize=8.9cm \epsfclipon
    \epsfbox{8399.f3b}
    \hfill
    \epsfxsize=8.9cm \epsfclipon
    \epsfbox{8399.f3e}
    \epsfxsize=8.9cm \epsfclipon
    \epsfbox{8399.f3c}
    \hfill
    \epsfxsize=8.9cm \epsfclipon
    \epsfbox{8399.f3f}
    \epsfxsize=18cm \epsfclipon
    \epsfbox{8399.f3g}
    \epsfxsize=18cm \epsfclipon
    \epsfbox{8399.f3h}    
    \caption{\label{fig:egretsim}
    Reconstruction of an EGRET $>100$ MeV template which was adjusted 
    to the expected 1.8 \MeV\ intensity by means of a model fit.
    In the profiles, the model is shown as solid, the ME reconstruction 
    as dotted, the RL reconstruction as dashed, and the MREM ($s=3.0$) 
    reconstruction as dashed-dotted line.
    } 
  \end{minipage}
\end{figure*}

Figure \ref{fig:egretsim} presents the results of the simulations based on 
the EGRET $>100$ MeV all-sky map.
This map shows a ridge of diffuse emission along the galactic plane 
with a notably intensity enhancement towards the inner Galaxy, 
some prominent galactic point sources, some localised emission regions,
and some extragalactic point sources.
The intensity level of the EGRET $>100$ MeV all-sky map has been adjusted 
to a plausible 1.8 \MeV\ intensity level by fitting the map to COMPTEL 
1.8 \MeV\ \gray\ line data, resulting in a 1.809 \MeV\ line flux from 
the inner radian ($|b|<20\degr$) of $\sim3\times10^{-4}$ \frad.
This adjustment pushed most of the point sources in the EGRET map 
below the sensitivity limit of COMPTEL at 1.8 \MeV, leaving Vela 
with $3\times10^{-5}$ \funit\ and Geminga with $1\times10^{-5}$ \funit\ 
as the most prominent objects.

Indeed, the only point source recovered in the ME and RL 
reconstructions is Vela ($l=264\degr$, $b=-3\degr$), while only a small 
hint of \gray\ emission is seen at the position of Geminga
($l=195\degr$, $b=-4\degr$).
Similar to the exponential disk simulation, `hot spots' and emission 
gaps appear along the galactic plane which only occasionally coincide 
with localised features in the model map.
Such coincidences are found e.g.~at $l\approx80\degr$ (Cygnus) or at 
$l\approx-45\degr$.
However, localised emission features are also found in the 
exponential disk simulations at these positions 
(cf.~Fig.~\ref{fig:expsim}) where no such features are present in the 
model.
Since only the mock dataset of the (dominant) instrumental background 
component is common to both simulations, it is very suggestive that 
the observed features are at least partially due to positive statistical 
fluctuations of the background data.
Other localised emission features in the EGRET map, like the spots at 
$l\approx20\degr$, $l\approx-18\degr$, or $l\approx-75\degr$ (Carina), 
coincide with negative statistical fluctuations of the background 
component and annihilate; consequently no feature is seen in the 
reconstructions at these positions.
Additionally, artificial `hot spots' appear in the ME and RL maps where 
no such features are present in the model.
Examples are the strong feature at $l\approx115\degr$ and the spur 
towards negative latitudes at $l\approx-15\degr$.
Again, these features can also be percepted in the exponential disk 
simulations, confirming that they arise from the statistical noise of 
the background sample.

In contrast to ME and RL, MREM again provides much smoother 
reconstructions of the data, avoiding most of the artifacts.
In the $s=2.5$ run, only the most prominent artifacts are visible 
(e.g.~the spot at $l\approx115\degr$), but many of the real localised 
features are recovered ($l\approx80\degr$, $l\approx135\degr$,
$l\approx-45\degr$, and Vela).
Increasing the requirement for the significance of the emission 
structures to $s=3.0$ removes the remaining artifacts, but eliminates 
also most of the localised emission features.
Nevertheless, weak hints for Vela and the $l\approx135\degr$ source 
are still present in the sky map.
These hints disappear when $s$ is increased to $3.5$.
Again, the latitude extent of the reconstructions is slightly higher 
than that of the models due to the combined result of the 
instrument's angular resolution of only $4\degr$ (FWHM) together with 
a low signal to noise ratio.
Yet, the extended diffuse emission above the galactic centre is still 
recovered in the maps.

The residual analysis of the MREM $s=3.0$ reconstruction reveals only 
weak emission at the position of the localised features, indicating that 
they are not very significant (cf.~Fig.~\ref{fig:egretres}).
The most prominent residuals are found at the position of Vela 
and at $(l,b)=(81\degr, -16\degr)$, with likelihood ratios of
$-2 \ln \lambda = 16.6$ and $14.3$, respectively.
While the first residual corresponds to a real source in the EGRET 
map, the second one is a clear background fluctuation.
If the existence of the Vela source would not be known a priori, 
the likelihood ratio of $16.6$ converts to a detection significance of 
$3.3\sigma$ (3 d.o.f.).
Taking into account the number of trials made in the point source 
search, this value can not be interpreted as a significant 
detection.
However, if the Vela source is considered as known object, the 
likelihood ratio converts to a $4.1\sigma$ detection significance 
(1 d.o.f.).
The major objective of an 1.809 \MeV\ all-sky map, however, is the 
discovery of unknown objects, hence it is desirable that the Vela 
source is not recovered in the reconstruction.
Otherwise, as demonstrated by the ME and RL or the MREM ($s=2.5$) 
reconstructions, artifacts will also enter the reconstruction, making 
the interpretation of the sky map difficult.

\begin{figure*}
  \setlength{\unitlength}{1cm}
  \begin{minipage}[t]{18cm}
    \epsfxsize=18cm \epsfclipon
    \epsfbox{8399.f4}
    \caption{\label{fig:egretres}
    Residual maximum likelihood ratio maps of the EGRET $>100$ MeV all-sky 
    map simulations.
    Contour levels like in Fig.~\ref{fig:expres}.
    From top to bottom the panels show 
    (a) the residuals of the background sample only (i.e.~the noise), 
    (b) the residuals of iteration 6 of the ME reconstruction,
    (c) the residuals of iteration 6 of the RL reconstruction,
    and (d) the residuals of the MREM ($s=3.0$) reconstruction.
    } 
  \end{minipage}
\end{figure*}

\begin{figure}
  \setlength{\unitlength}{1cm}
  \begin{minipage}[t]{8.8cm}
    \epsfxsize=8.8cm \epsfclipon
    \epsfbox{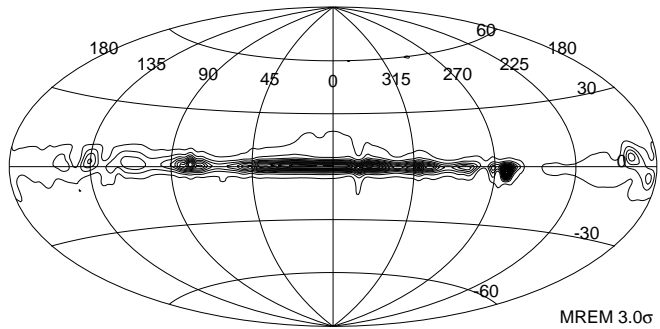}
    \caption{\label{fig:egret5}
    MREM ($s=3.0$) reconstruction of an EGRET $>100$ MeV template 
    with an intensity that is 5 times larger than the 1.809 \MeV\ 
    intensity.
    } 
  \end{minipage}
\end{figure}

To illustrate that MREM indeed recovers point sources if they are 
significant, we performed an additional simulation where we increased 
the intensity of the EGRET $>100$ MeV template by a factor of 5 
with respect to the 1.809 \MeV\ intensity.
This corresponds to an increase of a factor of 5 in the 
signal-to-noise ratio, which is equivalent to a sensitivity 
enhancement of the same magnitude.
The resulting MREM reconstruction is shown in Fig.~\ref{fig:egret5} 
for $s=3.0$.
As expected, much more structure along the galactic plane is now
recovered.
Prominent point sources, such as Vela, Geminga, or the Crab, and localised 
emission features, e.g.~in Cygnus and Carina, are now clearly visible.
This demonstrates that the absence of these features in the MREM map 
derived for the 1.809 \MeV\ intensity level (Fig.~\ref{fig:egretsim}) 
relates to their significance.

\section{The COMPTEL 1.8 MeV sky}
\label{sec:comptel}

The MREM algorithm is now applied to real COMPTEL 1.8 \MeV\ data, 
taken during observation periods 0.1 - 522.5 (May 1991 to June 1996).
The instrumental background component was estimated using contemporaneous 
data at adjacent energies, following the procedure described in 
Kn\"odlseder et al.~(1996).
The time-variability of the instrumental background was taken into 
account by determination of the background model on single observation 
basis, and by its proper relative normalisation using the activation history 
of major background components (Oberlack 1997).
Since the absolute normalisation as well as the $\phibar$ distribution 
of the background model are only weakly constrained, we added an 
additional step where we determine both by all-sky model fitting 
(maximum likelihood optimisation).
For this purpose we fitted the instrumental background model together 
with a template for the 1.809 \MeV\ intensity distribution to the COMPTEL 
1.8 \MeV\ data, where we determined independent scaling factors for all 
$\phibar$ layers of the background model as well as a global scaling 
factor for the 1.8 \MeV\ intensity template.
This procedure provides an estimate of the total 1.809 \MeV\ sky 
flux and an improved estimate of the instrumental background
component, which is then used for image reconstruction.
For the 1.809 \MeV\ template we used the 53 GHz free-free emission map 
derived from COBE/DMR data (Bennett et al.~1992) which was found to 
provide the best description of the COMPTEL 1.8 \MeV\ data in a recent 
study using a wide variety of models (Kn\"odlseder et al.~1999).
An alternative method for deriving an instrumental background model 
for the analysis of 1.8 \MeV\ \gray\ line data is described in Bloemen et 
al.~(1999b).

\begin{figure*}
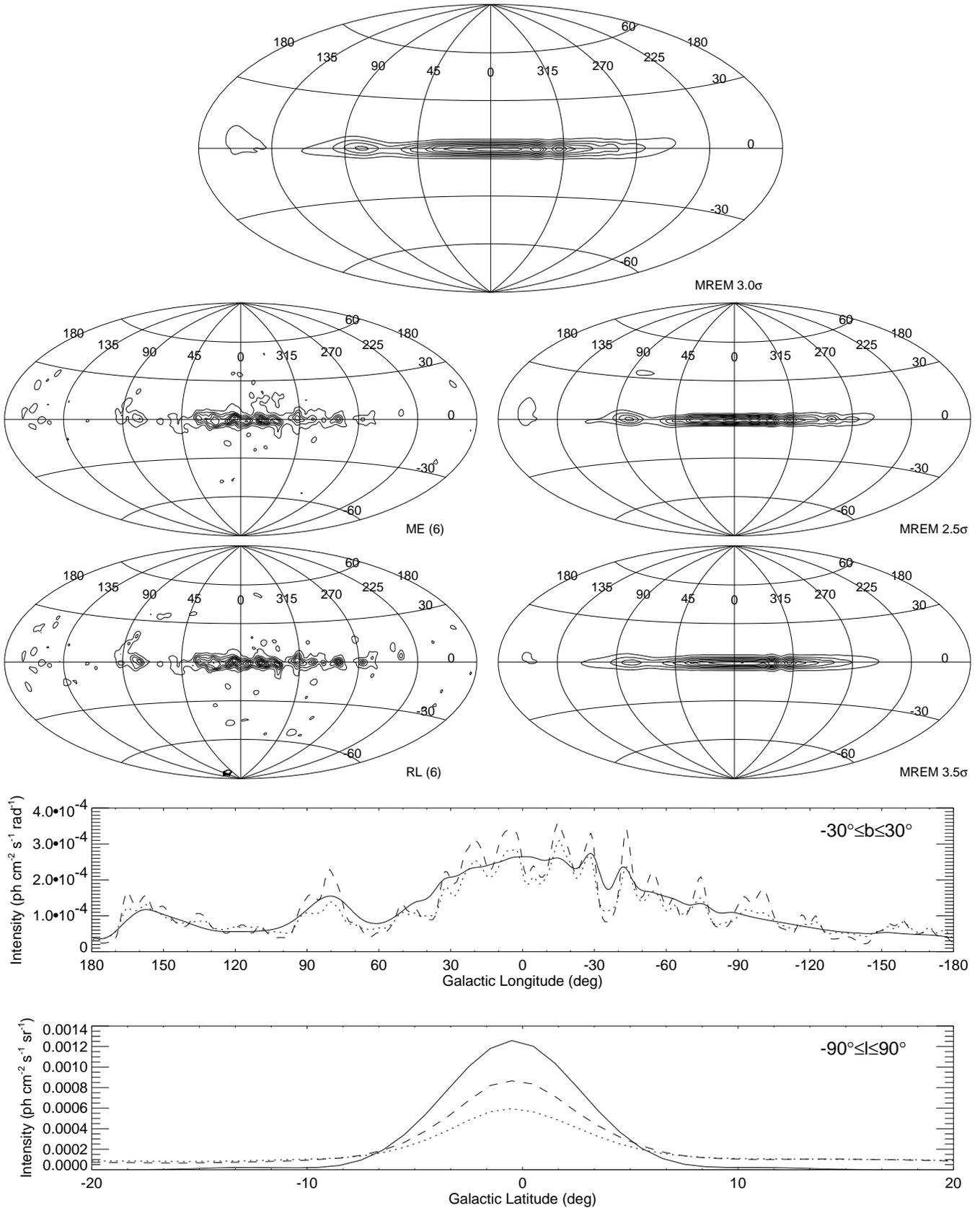

  \setlength{\unitlength}{1cm}
  \begin{minipage}[t]{18cm}
    \centerline{
    \epsfxsize=11cm \epsfclipon
    \epsfbox{8399.f6d}}
    \epsfxsize=8.9cm \epsfclipon
    \epsfbox{8399.f6a}
    \hfill
    \epsfxsize=8.9cm \epsfclipon
    \epsfbox{8399.f6c}
    \epsfxsize=8.9cm \epsfclipon
    \epsfbox{8399.f6b}
    \hfill
    \epsfxsize=8.9cm \epsfclipon
    \epsfbox{8399.f6e}
    \epsfxsize=18cm \epsfclipon
    \epsfbox{8399.f6f}
    \epsfxsize=18cm \epsfclipon
    \epsfbox{8399.f6g}
    \caption{\label{fig:realdata}
    COMPTEL 1.8 \MeV\ all-sky maps derived using the maximum entropy 
    method (ME), the Richardson-Lucy algorithm (RL), and the 
    Multiresolution Expectation Maximization algorithm (MREM).
    In the profiles, the MREM ($s=3.0$) reconstruction is shown as 
    solid, the ME reconstruction as dotted, and the RL reconstruction 
    as dashed line.
    } 
  \end{minipage}
\end{figure*}

Figure \ref{fig:realdata} shows the COMPTEL 1.809 \MeV\ \gray\ line 
all-sky maps that are obtained by the different reconstruction 
algorithms.
The maximum entropy and Richardson-Lucy maps are similar to those 
presented in previous work (Diehl et al.~1995; Oberlack et al.~1996; 
Kn\"odlseder et al.~1996; Oberlack 1997; Bloemen et al.~1999b) with minor 
differences being due to differences in the analysed data volume or the 
employed background modelling procedure.
The most distinct feature in these maps is emission along the ridge 
of the galactic plane.
Again we see the lumpiness that our above simulations also show for 
these methods, indicating overfitting of the data.
According to the discussion above we cannot decide from the sky maps 
alone which of the lumps may correspond to real emission and 
which are artifacts due to the background noise.

MREM avoids this confusion, suppressing efficiently the noise 
components of the image.
The reconstructed intensity profiles are characterised by a notable 
asymmetry with respect to the galactic centre and some localised emission 
features.
The most prominent of these features is located in the Cygnus region 
around $(l,b)\approx(80\degr,0\degr)$ where a bright extended emission 
spot is clearly separated from the inner galactic ridge emission by a 
bridge of relatively low 1.8 \MeV\ intensity.
The same feature is also seen in the ME and RL maps where it 
obeys a much more complex structure.
The MREM reconstructions suggest that most of this structure is not 
individually significant, and could as well be more diffuse or located 
differently.
We therefore safely may extract the fact of significant Cygnus 
region emission, separated from the inner galactic ridge.

\begin{figure*}
  \setlength{\unitlength}{1cm}
  \begin{minipage}[t]{18cm}
    \epsfxsize=18cm \epsfclipon
    \epsfbox{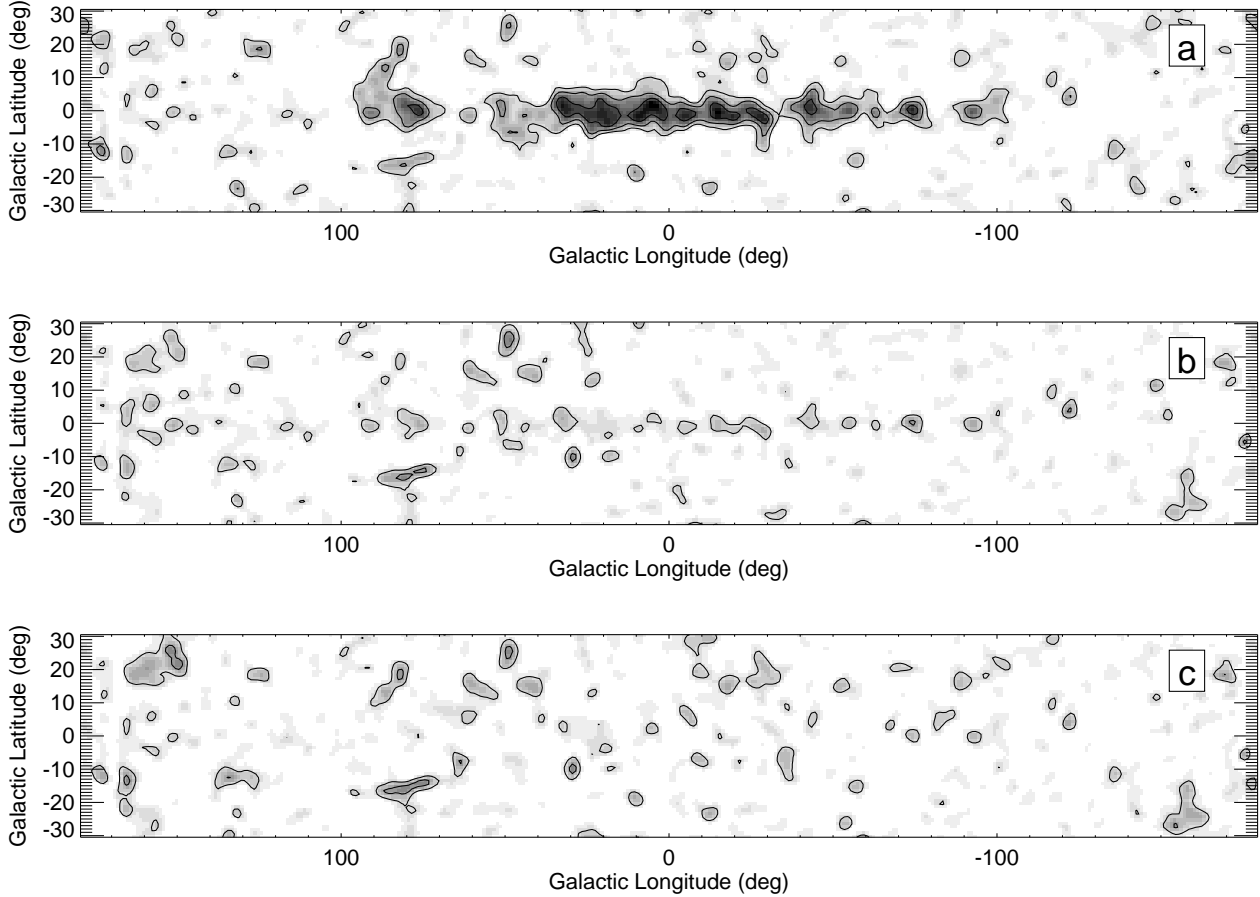}
    \caption{\label{fig:realdatares}
    Residual maximum likelihood ratio maps of the COMPTEL 1.8 \MeV\ 
    all-sky maps.
    Contour levels like in Fig.~\ref{fig:expres}.
    From top to bottom the panels show 
    (a) the residuals of iteration 6 of the ME reconstruction,
    (b) the residuals of iteration 6 of the RL reconstruction, and
    (c) the residuals of the MREM ($s=3.0$) reconstruction.
    } 
  \end{minipage}
\end{figure*}

This inner ridge appears smooth on the MREM image, yet also here 
reveals a pronounced asymmetry with respect to the galactic centre.
While at positive longitudes the intensity drops steeply from 
$l\approx30\degr$ to $l\approx50\degr$, the 1.8 \MeV\ emission extends 
continuously to $l\approx240\degr$ at negative longitudes.
Along the ridge the MREM reconstructions reveal only little 
structure.
From the many `hot spots' seen in the ME and RL reconstructions, only 
the most prominent ones are still perceptible in the MREM image obtained 
for $s=2.5$.
Increasing the significance level to $s=3.0$ removes most of them, 
keeping only two emission spots at $l=317\degr$ and $332\degr$ which 
are separated by a weak emission gap at $l=324\degr$.
This is the most persistent structure along the inner galactic plane 
ridge which is still clearly visible for a significance level of 
$s=3.5$.
It is also very pronounced in the ME and RL maps.
Additional hints for weak excess emission are found in the 
longitude profile at $l=21\degr, 30\degr, 44\degr, 286\degr$, and 
$345\degr$, but they disappear if $s$ is increased to $3.5$.
Obviously, the significance of these excesses is close to the 
sensitivity limit of COMPTEL, and the assessment of their reality 
needs more dedicated studies.

The distinct emission gap which separates two localised emission 
regions at $l=266\degr$ (Vela) and $l=286\degr$ (Carina) in the ME and 
RL maps is not seen in the MREM reconstructions.
Yet, a weak intensity dip is found in the $s=2.5$ and $3.0$ MREM maps 
at this location, indicating that some structure may indeed be present.
The prominent `hot spot' towards the galactic centre, clearly visible 
in the ME and RL maps at $l=4\degr$, is only present in the MREM map 
for $s=2.5$, but disappears for higher significance levels.
Apparently, the data are also consistent with a smooth emission 
profile in this region.

Comparison of the 1.8 \MeV\ maps with the EGRET map simulations 
indicates that the 1.809 \MeV\ emission is very confined to the 
galactic plane.
In particular, there is no hint for an extended emission component 
similar to that seen in the EGRET map above the inner Galaxy.
Indeed, model fitting using exponential disk models revealed a small
scale height of $z_{0}=90$ pc for the galactic \al\ distribution 
(Kn\"odlseder 1997).
Comparison of the 1.8 \MeV\ maps with the exponential disk 
simulations (for which a scale height of $z_{0}=90$ pc was assumed) 
confirms this result.
In particular, the width of the MREM ($s=3.0$) 1.8 \MeV\ latitude profile 
($7.5\degr$ FWHM) is even smaller than that obtained for the exponential 
disk simulation ($8.8\degr$ FWHM), indicating that the scale height of 
the \al\ distribution may be even below 90 pc.

Near the anticentre, all maps of Fig.~\ref{fig:realdata} show 
indications for extended 1.8 \MeV\ \gray\ line emission.
In the ME and RL reconstructions, weak emission spots are spread over 
a region extending from $125\degr - 170\degr$ in galactic 
longitude and from $-20\degr - 30\degr$ in galactic latitude.
The MREM algorithm combines these spots to a more concentrated 
emission structure, roughly located at $l\sim160\degr$ with an angular 
extent of $\sim20\degr$.
This again illustrates that the spots in the ME and RL images are not 
significant for themselves, but when combined they provide a 
significant 1.809 \MeV\ emission feature.

Residual maximum likelihood ratio maps of the COMPTEL 1.8 \MeV\ all-sky 
maps are compiled in Fig.~\ref{fig:realdatares}.
The ME reconstruction shows significant residual emission along the 
galactic plane which is strongly correlated to the reconstructed sky 
intensity distribution.
The intensity profiles in Fig.~\ref{fig:realdata} illustrate that
iteration 6 of the ME reconstruction considerable underestimates 1.8 
\MeV\ intensities with respect to RL and MREM.
For higher ME iterations, this underestimation disappears as the 
maximum entropy reconstruction approaches the maximum likelihood 
solution.
Yet, the diffuse intensity distribution breaks up into nearly 
isolated point sources for late iterations due to overfit of the 
data (Kn\"odlseder et al.~1996).
Therefore we typically present COMPTEL ME images and longitude profiles 
from `early' iterations in order not to emphasise artificial 
structures, while `late' iterations are used to derive 1.809 \MeV\ fluxes 
and latitude profiles in order to recover the correct flux values 
(Diehl et al.~1995; Oberlack et al.~1996).
Alternatively, intensity distributions for `late' iterations have been 
smoothed to the instrumental resolution for image presentation to reduce 
the artificial lumpiness (Oberlack 1997; Strong et al.~1999).

Also the RL reconstruction shows residuals that are correlated with the 
galactic plane, although they are much smaller than for ME.
Yet there are regions where almost no residuals are found, in 
particular at negative galactic longitudes ($l<0\degr$) above and 
below the galactic plane.
Comparison with the simulations suggests that the lack of residuals 
is again due to overfit of the data.
In contrast, the MREM $(s=3.0)$ reconstruction provides residuals 
that appear uncorrelated with the galactic plane.
This clearly illustrates that the MREM map presents a statistical 
satisfactory description of COMPTEL 1.8 \MeV\ data.

\section{Conclusions}
\label{sec:conclusions}

An alternative imaging method for COMPTEL \gray\ data is presented, 
using a newly developed multiresolution reconstruction algorithm based 
on wavelets.
The maximum entropy and Richardson-Lucy algorithms, which have been 
used previously for COMPTEL image reconstruction, are very sensitive to 
statistical noise in the data, leading to image lumpiness and `hot spots'
in the reconstructed intensity maps.
In particular, artificial `hot spots' are indistinguishable from real 
point sources on basis of the sky maps alone, requiring an 
substantial additional analysis effort to assess their reality.
We present the resulting reconstructed image of the 1.809 \MeV\ sky 
as an alternative view, complementing previously presented images from 
the other methods, and pointing out their limitations.
In particular, we caution to overinterpret structure in the ME and RL 
sky maps when modelling the galactic \al\ emission for other studies
(e.g.~Lentz et al.~1998).

Applying our new algorithm to COMPTEL data largely reduces or even 
removes artificial `hot spots' and image lumpiness, depending on the 
selected significance requirement $s$.
Simulations indicate that $s=3.0$ seems to provide a reasonable 
choice for the reconstruction: while artifacts are mainly removed from 
the image, hints for weak ($3-4\sigma$) point sources are still present.
Nevertheless, it should be clear that the MREM sky map obtained in 
this work not necessarily provides a realistic view of the 1.809 
\MeV\ sky.
The real 1.809 \MeV\ intensity profile is probably much more confined 
to the galactic plane than the emission in the MREM sky map, but with an 
angular resolution of $4\degr$ (FWHM), COMPTEL is not capable of resolving 
this confinement.
The 1.809 \MeV\ emission along the galactic plane may be much more 
structured than shown in the MREM map, but the sensitivity of COMPTEL 
is not sufficient to map this structure.
In this sense, MREM provides a more reliable image of the 1.809 \MeV\ 
\gray\ sky with respect to ME and RL since it does not show emission 
structures for which there is no strong evidence.
Weak emission features which are close to the sensitivity limit of 
COMPTEL may however be suppressed in the MREM maps.
Therefore ME and RL maps are used as complementary analysis tools.

\begin{acknowledgements}

JK is supported by the European Community through grant number
ERBFMBICT 950387.
The COMPTEL project is supported by the German government through
DARA grant 50 QV 90968, by NASA under contract NAS5-26645, and by
the Netherlands Organisation for Scientific Research NWO.
This research has made use of data obtained through the High Energy 
Astrophysics Science Archive Research Center Online Service, provided 
by the NASA/Goddard Space Flight Center.

\end{acknowledgements}


\end{document}